\documentclass[runningheads]{svmult}
\usepackage{natbib,epsfig}
\usepackage{makeidx}   
\usepackage{graphicx}  
\usepackage{subeqnar}  
\usepackage{multicol}  
\usepackage{cropmark} 
\usepackage{physprbb}  
\newcommand{\greeksym}[1]{{\usefont{U}{psy}{m}{n}#1}}
\newcommand{\uLambda}{\mbox{\greeksym{L}}}
\newcommand{\uOmega}{\mbox{\greeksym{W}}}

\begin{document}
\title*{Is the \uLambda CDM Model Consistent with \protect\newline
	Observations of Large-Scale Structure?}
\toctitle{Is the \uLambda CDM Model Consistent with Observations 
	\protect\newline of Large-Scale Structure?}
%
%
%
\titlerunning{Is \uLambda CDM Consistent with Large-Scale Structure?}
%
\author{Eric Gawiser
}
%
\authorrunning{Eric Gawiser}
%
%
\institute{Center for Astrophysics and Space Sciences, U. C. San Diego,
La Jolla, CA  92093}

\maketitle              

\begin{abstract}

The claim that large-scale structure data independently prefers 
the \uLambda~Cold Dark Matter 
model is a myth.  However, an updated compilation 
of large-scale structure observations cannot rule out \uLambda CDM
at 95\% confidence.  We explore the possibility of improving the model  
by adding Hot Dark Matter but the fit becomes worse;  this 
allows us to set limits on the neutrino mass.

\end{abstract}

\section{The Recent Past}

It has been known for several years that the 
Standard Cold Dark Matter model, 
(SCDM, \uOmega$_m=1, h=0.5, n=1$)\footnote{\uOmega$_x$ refers to the 
fraction of critical energy density present in component $x$; $m$  
represents all forms of matter, $\nu$ signifies massive neutrinos, and
\uLambda~stands for the cosmological constant.  
Hubble's constant is given by $H_0 = 100 h$km/s/Mpc.  $n$ refers
to the power-law index of the primordial power spectrum.}  
 could not simultaneously 
agree with COBE and large-scale structure observations.  Viable alternatives 
included Tilted Cold Dark Matter (TCDM, \uOmega$_m=1, h=0.5, n=0.8$), 
Cold + Hot Dark Matter (CHDM, \uOmega$_m=1$, \uOmega$_\nu=0.2, h=0.5, n=1$), 
Open Cold Dark Matter (OCDM, \uOmega$_m=0.3, h=0.7, n=1$), and
\uLambda CDM (\uOmega$_m=0.3$, \uOmega$_\Lambda=0.7, h=0.7, n=1$).

Gawiser \& Silk\cite{gawisers98} performed a 
quantitative comparison of these models with a compilation of 
CMB anisotropy and large-scale structure data and found that  
CHDM was the most successful model.
\uLambda CDM, OCDM, and TCDM were inconsistent with the 
data at 99\% confidence.  
The discrimination between models 
came primarily from large-scale structure 
data, particularly the APM galaxy power spectrum, and 
different values of the shape parameter 
($\Gamma$ = \uOmega$_m h$ in $n=1$ CDM models) 
were preferred on large and small scales.

\section{The Present}

Despite the recent success of CHDM, 
\uOmega$_m=1$ is not viable given current measurements of the cluster 
baryon fraction~\cite{evrard97} and the abundance of baryons inferred 
from measurements of deuterium in Lyman limit systems~\cite{burlest98}.  
Additionally, the high abundance of massive clusters at $z>0.5$
can only be reconciled with \uOmega$_m=1$ in the case of
 non-gaussian primordial density fluctuations~\cite{robinsongs00}. 

The overwhelming current evidence for a low matter density leaves us 
with only two of the previous models, \uLambda CDM and OCDM.  
The redshift-luminosity distance relationship observed for 
Type Ia supernovae data prefers \uLambda CDM to OCDM at better than 99\% 
confidence~\cite{perlmutteretal98, garnavichetal98a}, leaving us 
with a clear favorite of these ``direct'' cosmological tests and 
promoting \uLambda CDM to the status of the ``standard model'' of cosmology.

Should we be concerned that the same \uLambda CDM model was 
ruled out by \cite{gawisers98} at 99\% confidence?  Certainly.  However, given the history of systematic errors in astronomy it may well turn out that one 
or more of the aforementioned observations is flawed.  Given the confluence 
of evidence for a low value of \uOmega$_m$, the most likely culprit would be 
the observations of large-scale structure that drove the analysis of 
\cite{gawisers98} to favor CHDM.  The strongest discriminator was 
the real-space galaxy power spectrum inferred from the APM galaxy angular 
correlation function, and this has recently been carefully re-analyzed by 
Eisenstein \& Zaldarriaga~\cite{eisensteinz99}.  They find increased 
uncertainties on large spatial scales, and adopting their 
re-analysis leads to significant improvement in the agreement of
\uLambda CDM and OCDM with large-scale structure observations.
However, this agreement remains imperfect;  
CHDM is still preferred by large-scale structure data but less significantly.

This stands in marked contrast to a {\it myth} propagated by 
some cosmologists that large-scale structure provides independent evidence in 
favor of \uLambda CDM.  This myth appears to be caused by remembering
 that \uLambda CDM was 
preferred to Standard CDM by large-scale structure observations, 
knowing that \uLambda CDM has now been promoted to ``standard model,''  
and concluding that \uLambda CDM must be favored by 
large-scale structure data over all other models.  This is not correct.

\subsection{Available Large-Scale Structure Data}

Our large-scale structure data compilation includes 
the real-space galaxy power spectrum derived from the APM 
angular correlation function by \cite{eisensteinz99} and  
redshift-space power spectra from 
SSRS2+CfA2~\cite{dacostaetal94}, 
LCRS~\cite{linetal96}, 
PSCz~\cite{sutherlandetal99},
and APM Clusters~\cite{tadrosed98}.   
We also use measurements of the amplitude of the 
dark matter power spectrum at 8$h^{-1}$Mpc ($\sigma_8$) from the abundance of 
rich galaxy clusters at $z\simeq 0$~\cite{vianal96} 
and $z \simeq 0.3$~\cite{bahcallfc97} and 
at larger scales from peculiar velocity fields~\cite{kolattd97}.  

\begin{figure}[htb]
\begin{center}
\includegraphics[width=.99\textwidth]{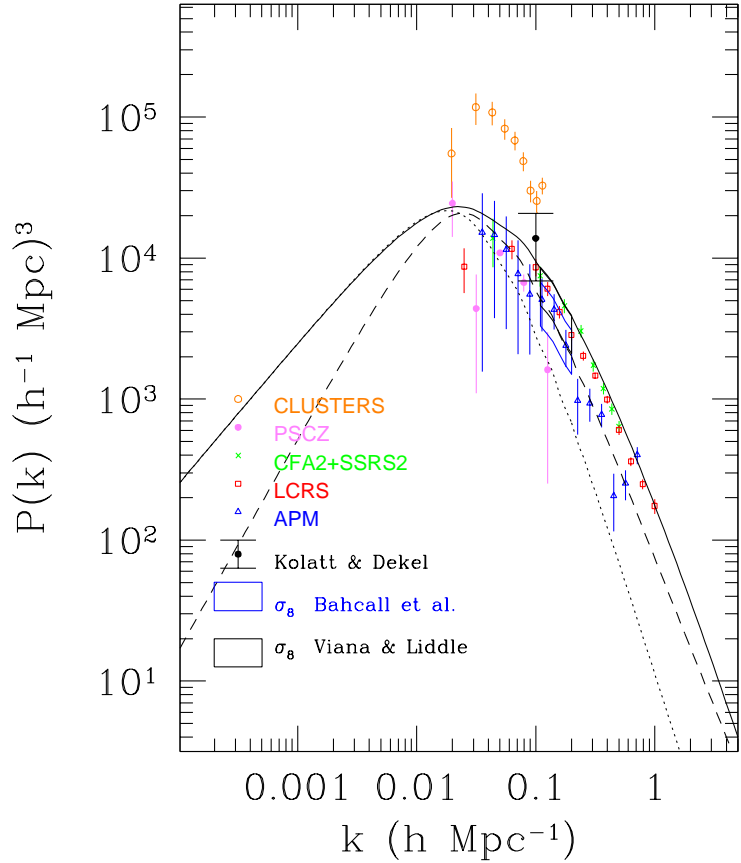}
\end{center}
\caption[]{ Comparison of large-scale structure
 data compilation with \uLambda CDM (solid),  
\uLambda CHDM with \uOmega$_\nu = 0.1$ and $n=1.0$ (dotted), and 
\uLambda CHDM with \uOmega$_\nu = 0.1$ and $n=1.5$ (dashed).  }
\label{lss_lchdm_all}
\end{figure}

Figure \ref{lss_lchdm_all} shows our data compilation versus 
the predictions for the matter power spectrum from the \uLambda CDM 
model and two versions of \uLambda CHDM with massive neutrinos 
generating \uOmega$_\nu=0.1$, one of which has a tilted ($n=1.5$) primordial 
power spectrum.  For quantitative results, we corrected for 
scale-independent galaxy bias, 
redshift distortions and non-linear evolution in the manner described by 
\cite{gawisers98, gawiser99}, 
but those corrections are not shown here.  Uncertainty 
in those corrections on non-linear scales at $k > 0.2$ 
leads us to restrict our analysis to larger scales.

\subsection{Cosmological Limits on Neutrino Masses}

Since the agreement between \uLambda CDM and the large-scale 
structure data is less than perfect, we investigate whether adding 
Hot Dark Matter will improve the fit (for a fuller 
discussion, see \cite{gawiser99,gawiser00}).  The opposite occurs; 
Fig.~\ref{lss_lchdm_all} shows that the \uLambda CHDM model with 
\uOmega$_\nu = 0.1, n=1$ is in 
serious disagreement with our large-scale structure data.  The 
diffusion of relativistic neutrinos in the early universe predicts 
a level of small-scale perturbations significantly 
lower than that observed.  However, 
tilting the primordial power spectrum to $n=1.5$ essentially 
resolves this problem.  
For large-scale structure data alone, this illustrates a degeneracy between 
the possibility of neutrino mass and the uncertain nature of the primordial 
power spectrum.  
Using all available CMB anisotropy data as well 
as our large-scale structure data compilation 
helps to differentiate 
between variations in the primordial power spectrum and the reduction in 
small-scale power caused by massive neutrinos.

\begin{figure}[htb]
\begin{center}
\includegraphics[width=.99\textwidth]{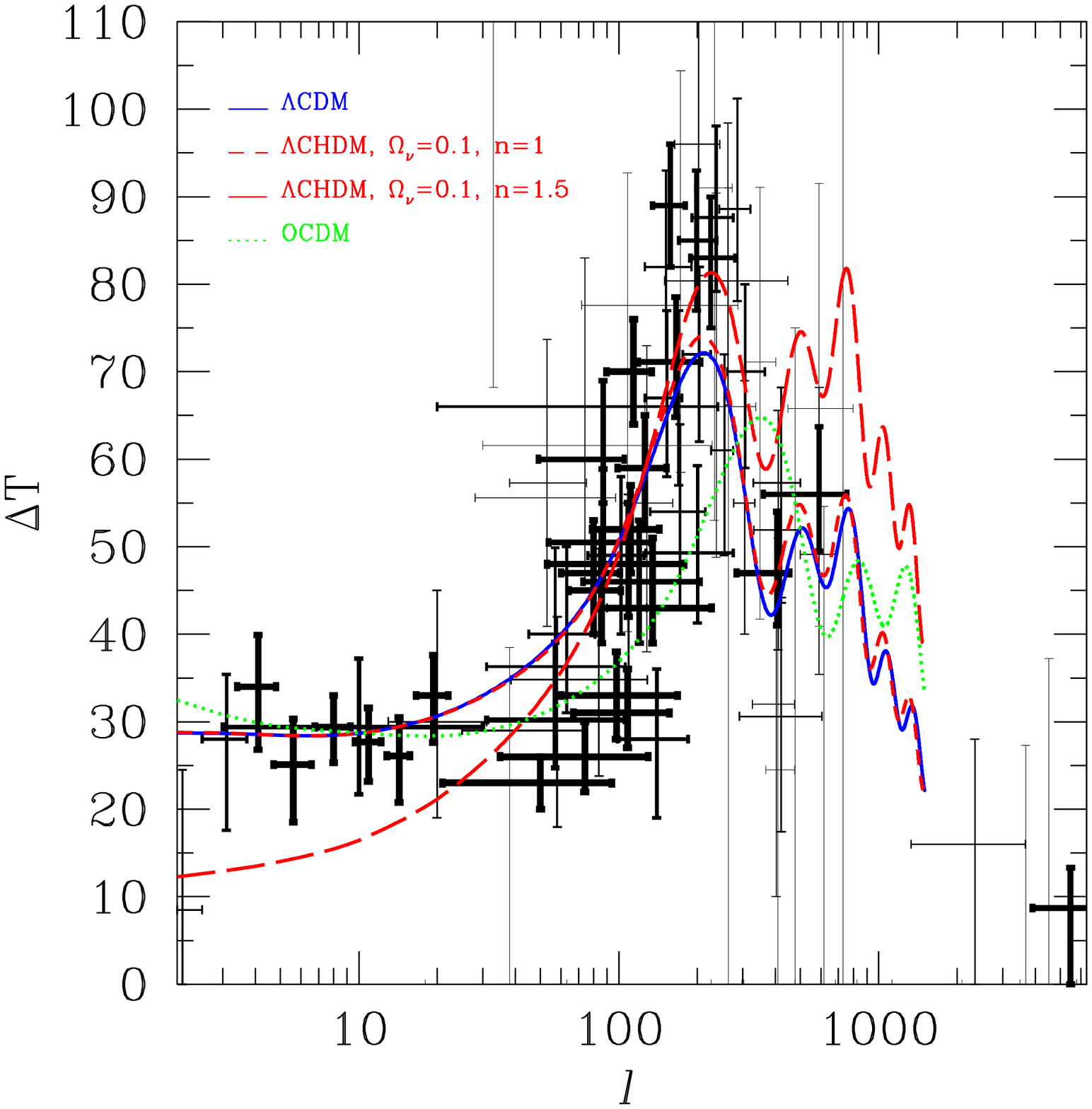}
\end{center}
\caption[]{ CMB anisotropy observations.  The thickness of line used 
for each point is proportional to the inverse of its variance; this prevents 
highly uncertain points from dominating the plot.  Theoretical predictions 
are shown for \uLambda CDM (solid blue),  
\uLambda CHDM with \uOmega$_\nu = 0.1$ and $n=1.0$ (short-dashed red), 
\uLambda CHDM with \uOmega$_\nu = 0.1$ and $n=1.5$ (long-dashed red), 
and OCDM (dotted green). }
\label{dT_lchdm}
\end{figure}

Figure \ref{dT_lchdm} illustrates the complementary nature of CMB 
anisotropy data.  The Open CDM model is ruled out by the location 
of the first peak caused by acoustic oscillations, which instead indicates 
a flat geometry for the universe.  The \uLambda CHDM model with $n=1.5$ 
predicts too much anisotropy at high multipoles (small angles), whereas 
the presence of massive neutrinos is indistinguishable for $n=1$.  
Thus the version of \uLambda CHDM that agreed well with large-scale structure 
data is ruled out by the CMB, and vice versa.

We have assumed either a Harrison-Zel'dovich (scale-invariant, $P_p(k)=Ak$) 
or a scale-free ($P_p(k)=Ak^n$) primordial power spectrum. 
If \uLambda CDM is right and $n=1$, then 
\uOmega$_\nu \leq 0.05$, and $m_\nu \leq 2$~eV.  
If $n$ can vary, 
\uOmega$_\nu \leq 0.1$, and $m_\nu \leq 4$~eV 
(limits are at 95\% confidence).  
More freedom in 
the primordial power spectrum makes it nearly impossible to limit the 
neutrino mass.

\section{The Future}

Interesting physics can be probed by large-scale structure with forthcoming 
data from the 2-Degree Field survey (2dF) and the Sloan Digital Sky 
Survey (SDSS).  
The location of the peak in the matter power spectrum identifies the 
horizon size at matter-radiation equality ($z\sim 10^4$), 
and for pure CDM models this provides an 
independent measurement of $\Gamma =$ \uOmega$_m h$.  
Large-scale structure data at $k=$0.1--0.2 probe the 
primordial power spectrum 
on smaller scales than can be measured well with CMB anisotropy data 
before Planck; this provides a much-needed {\it test of inflation}.  
As discussed by Hu, Eisenstein, \& Tegmark~\cite{huet98}, 
the method utilized here can hope to 
measure (or constrain) neutrino masses down to 0.5 eV given 
data from the MAP satellite and SDSS.

\section{Conclusions}

\uLambda CDM is now the standard cosmological model 
but large-scale structure 
data does not prefer it over CHDM or OCDM.  Current large-scale structure 
data cannot rule out any of those models at 95\% confidence.

\uLambda CDM does not prefer the addition 
of a Hot Dark Matter component in the form of massive neutrinos. 
This leads to an upper limit on the 
mass of the most massive neutrino of 4~eV if a power-law primordial 
power spectrum is assumed.

Forthcoming observations of large-scale structure from 2dF and SDSS 
will probe the horizon size at matter-radiation equality, the 
primordial power spectrum on small scales, and neutrino masses.

\nopagebreak











%

\end{document}